# Dominant Kinetic Pathways of Graphene Growth in Chemical Vapor Deposition: The Role of Hydrogen

*Pai Li, Zhenyu Li,\* and Jinlong Yang*

Hefei National Laboratory for Physical Sciences at the Microscale, University of Science and Technology of China, Hefei, Anhui 230026, China

**ABSTRACT:** The most popular way to produce graphene nowadays is chemical vapor deposition, where, surprisingly, $H_2$ gas is routinely supplied even though it is a byproduct itself. In this study, by identifying dominant growing pathways via multiscale simulations, we unambiguously reveal the central role hydrogen played in graphene growth. Hydrogen can saturate the edges of a growing graphene island to some extent, depending on the $H_2$ pressure. Although graphene etching by hydrogen has been observed in experiment, hydrogen saturation actually stabilizes graphene edges by reducing the detachment rates of carbon-contained species. Such a new picture well explains some puzzling experimental observations and is also instrumental in growth protocol optimization for two-dimensional atomic crystal van der Waals epitaxy.

**Introduction**

Chemical vapor deposition (CVD) on a catalytic metal substrate has been widely recognized as the method of choice to produce high-quality graphene samples[1,2] in large quantities.[3] Copper is the most popular substrate used in graphene CVD growth and different recipes have been proposed to improve the sample quality and/or the growth rate.[4] Compared to other highly catalytic metal substrates with a large carbon affinity,[5–7] graphene growth on Cu surfaces is more complicated. Previous mechanism studies mainly focus on H-free carbon species[8–15] and much of the underlying atomistic mechanisms for CVD growth, which is typically supplied with both $CH_4$ and $H_2$, remains elusive.

In graphene CVD growth, $H_2$ partial pressure is a critical experimental parameter which can affect the sample morphology[16,17] and thickness.[18–20] Both hydrogen promoted graphene growth[21,22] and graphene etching by hydrogen[23–25] have been observed in experiment, which lead to controversial mechanisms[21,26] to describe the role hydrogen played in graphene growth. Generally, via hydrogenation/dehydrogenation reactions, hydrogen can change both surface species concentrations and graphene island edge configuration, which then determines the attachment/detachment dynamics of surface species to/from graphene edges. To understand such a complicated kinetic network, dominant kinetic pathways should be unambiguously identified via an extensive study of elementary atomic steps.

In this study, by combining first-principles calculations and effective kinetic Monte Carlo (KMC) simulations, we systematically investigate the mechanisms of graphene CVD growth on Cu substrate. Under most experimental conditions, H adatom is found to be the most abundant surface species, which determines the degree of H-saturation of graphene edges. Due to its extremely low concentration on the surface, detachment rate of a C-contained species instead of its attachment rate determines its net contribution to graphene growth. Since detachment from H-saturated edge sites is generally more difficult, hydrogen saturation stabilizes graphene edge in most experimental conditions, which explains why the $CH_4 \rightarrow C + 2H_2$ reaction can be promoted by adding extra $H_2$ molecules. Experimentally observed graphene etching by hydrogen is actually a result of the hydrogenation reaction of detached species, which shifts the chemical equilibrium to the etching end via desorption of hydrogenated species into the gas phase. Results reported here clarify the central role hydrogen played in graphene CVD growth, which should be instrumental in gaining more precise control of epitaxial graphene growth for various device applications.

**Computational Details**

Electronic structure calculations were carried out with the density functional theory (DFT) implemented in the Vienna ab initio simulation package (VASP).[27,28] The general gradient approximation (GGA) to exchange-correlation functional parameterized by Perdew, Burke and Ernzerhof (PBE)[29] was adopted with a DFT-D2 correction[30] to describe van der Waals (vDW) interactions. Projector augmented wave method[31] was used to describe the core-valence interaction. A kinetic energy cutoff of 400 eV was used for the plane wave basis set. The energy convergence threshold was set to $1 \times 10^{-5}$ eV, and atom positions were relaxed with the conjugate gradient method until the force on each atom was less than 0.02 eV/Å. The climbing image nudged-elastic-band (CI-NEB) method was used for transition state location.[32] Cu(111) surface was modeled using a four-layer slab which is separated with its neighboring images by a 15 Å vacuum layer. All atoms, except those in the bottom layer of the slab, which were fixed to their corresponding bulk structure, were fully relaxed during geometry optimizations. For Cu(111) surfaces with adsorbates, a 4 × 4 supercell was used in the two dimensions parallel to the surface. For substrate with graphene edges, a 7 × 4 supercell was used. Their corresponding Monkhorst−Pack k-point sampling grid[33] was 4 × 4 × 1 and 2 × 4 × 1, respectively.

Standard rejection-free KMC simulation approach, namely the BKL algorithm,[34,35] was used to study graphene growth on Cu substrate. The Cu(111) surface was projected into a rectangular lattice with specific neighboring-site rules. Periodic boundary condition was applied using large supercells up to 2000 × 2000. When focused on C-contained species, considering the high concentration and rapid diffusion rate of H adatoms on Cu(111), we adopted a mean field approximation (MFA) to hydrogenation and dehydrogenation reactions, which greatly speeded up KMC simulations by only recording the number of H adatoms and adjusting the hydrogenation/dehydrogenation rates accordingly. Test calculations demonstrated that results reported here are not affected by such an approximation. All KMC simulations were performed at 1300 K. KMC codes were written in the C/C++ language. More details about KMC simulations can be found in the Supporting Information.

**Results and Discussion**

To understand the growth mechanism, we first need to know what are the species existing on the surface. One important species is atomic H, which can be generated by dissociative adsorption of gas-phase $H_2$ molecules. Our DFT calculations predict a moderate activation barrier for both the $H_2$ dissociation and its reverse reactions (0.31 and 0.80 eV, respectively). Together with the high mobility of H adatom on Cu(111), these results suggest that the equilibrium between H adatoms and gas-phase $H_2$ molecules can be easily reached at the high temperatures during graphene growth. As a result, the gas phase acts as an effective reservoir to control the number of surface H adatoms, which leads to the following relationship between the surface H adatom concentration [H] and the $H_2$ partial pressure

$$[H] = 4.08 \times 10^{-4} \times \sqrt{P_{H_2}},$$

where [H] is in ML and $P_{H2}$ is in Torr (see Supporting Information for more details).

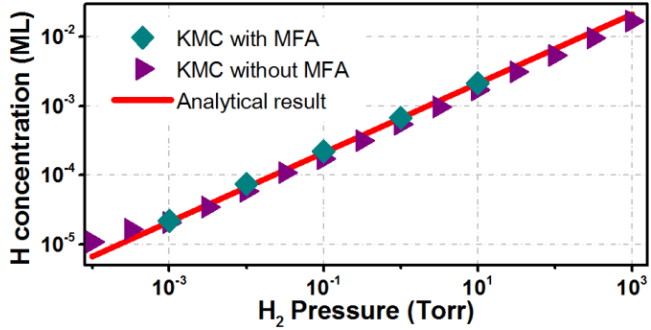

**Figure 1.** Concentration of H adatoms versus gas-phase $H_2$ pressure. KMC results with the mean field approximation (MFA) are obtained under a $CH_4$ partial pressure of 10 Torr.

For other species, considering that growth is a non-equilibrium process, we obtain their steady-state concentrations directly via KMC simulations instead of from formation energies under an equilibrium-state approximation.[36,37] Starting from $CH_4$ and $H_2$ decomposition, many hydrocarbon species can exist on the surface. Previous studies mainly focus on $CH_x$ species.[37,38] Motivated by our recent result that $C_2$ is the dominant feeding species in H-free graphene growth,[15] we also include $C_2H_x$ species in this study. All possible conversion processes among these species (H, C, CH, $CH_2$, $CH_3$, $CH_4$, $C_2$, $C_2H$, and $C_2H_2$) are considered in our KMC simulations. At the same time, the gas phase with specific $H_2$ and $CH_4$ partial pressures is considered as a reservoir, which provides $H_2$ and $CH_4$ for surface adsorption and accommodates $H_2$, $CH_4$, and $C_2H_2$ in desorption events. Test simulations with larger species, such as $C_3$ and $C_3H$, give similar results (Figure S6 in Supporting Information).

As shown in Figure 1, the H adatom concentrations from KMC simulation agree well with the analytic result, which confirms that it is mainly determined by the $H_2$ partial pressure. Concentration of H adatoms is usually much higher than those of C-contained species (Figure 2), since the dissociative adsorption of $CH_4$ on Cu(111) is known to be unfavorable.[38] In fact, even if the $H_2$ pressure is several orders of magnitude lower than the $CH_4$ pressure, the adsorption rate of $H_2$ can still be much larger (see Supporting Information for details). Among C-contained species, $C_2$ has the highest steady-state concentration in most cases. The dominance of $C_2$ becomes less significant with the increase of the $H_2$ partial pressure. Eventually, CH becomes the most abundant C-contained species under high $H_2$ pressures. Other C-contained species are generally less important than $C_2$ and/or CH.

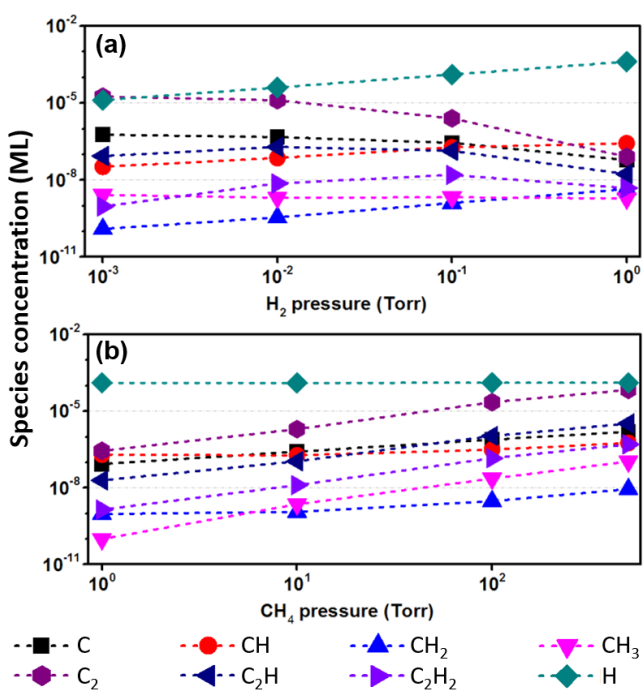

**Figure 2.** Initial concentrations of different species on the Cu surface for graphene growth under different $CH_4$ and $H_2$ partial pressures. (a) $CH_4$ pressure is fixed at 10 Torr and (b) $H_2$ pressure is fixed at 0.1 Torr.

In our KMC simulations, the non-equilibrium steady state can be reached typically in about $10^4$ seconds (Figure S4 in Supporting Information), which are expected to be much shorter than the time scale of graphene nucleation. Therefore, results obtained here give us the initial concentrations of different species during graphene CVD growth. They are generally different from concentrations obtained under an equilibrium-state approximation (details can be found in Supporting Information), which means that kinetics should be explicitly considered to unambiguously identify important surface species in graphene growth. Even if an effective equilibrium-state model can be constructed, kinetics-dependent artificial carbon/hydrogen chemical potentials should be used there to calculate formation energies.

Notice that, generally, activation barriers of surface reactions can be affected by coverage due to the adsorbate-adsorbate interaction.[39] Fortunately, in this study, we are usually in the dilute limit due to the low adsorption rate of $CH_4$ at the high temperature (about 1300 K) of graphene CVD growth. As shown in Figure 2, concentrations of carbon-contained species are always lower than $10^{-4}$ ML. Therefore, in most cases, the adsorbate-adsorbate interaction and thus the coverage effect on activation barriers can be safely neglected. However, as discussed below, in some cases such as hydrogen attachment/detachment, the coverage effect becomes important and it is thus explicitly considered in our KMC simulations.

After the surface concentrations of different species are obtained, the next thing should be determined is the graphene edge configuration. Under H-free growth conditions, graphene edges are passivated by surface Cu atoms, which prohibits edge reconstruction.[40] In CVD experiment, hydrogen saturation of edge dangling bonds provides a mechanism to compete with metal passivation. It is expected that graphene edge will be hydrogen-saturated under high $H_2$ pressures and metal-passivated under low $H_2$ pressures.[41] The question is if a homogeneous partially H-saturated edge can be formed or not under intermediate pressures. To answer this question, contribution from both enthalpy and entropy should be equally considered. A simple estimation of their relative importance is given below.

For a graphene edge with an H coverage λ, there are two possibilities. The edge can either be partially H-saturated forming a homogeneous phase or be segregated into fully H-saturated and metal-passivated phases. The free energy difference ($\Delta G$) between the homogeneous and inhomogeneous configurations is contributed by two parts

$$\Delta G = \Delta H - T\Delta S$$

The enthalpy contribution $\Delta H$ is calculated from the DFT energies of the metal-passivated phase ($E_0$), the homogenous partially H-saturated phase ($E_\lambda$), and the fully H-saturated phase ($E_1$). Since the enthalpy of the inhomogeneous configuration is simply an average of its two compositing phases,

$$\Delta H = E_\lambda - (1-\lambda)E_0 - \lambda E_1$$

In the entropy contribution, we only take the main part, i.e. configuration entropy, into account. It is about zero for the inhomogeneous configuration. Therefore, $\Delta S$ equals to the configuration entropy of the homogeneous system:

$$\Delta S = k_B \ln \frac{N!}{n!(N-n)!}$$

where $N$ is the number of total edge sites, $n=\lambda N$ is the number of H-saturated sites. Since $N$ is a very large number, according to the Stirling's formula

$$\ln X! \approx X(\ln X - 1)$$

we have

$$\Delta S = -[\lambda \ln \lambda + (1-\lambda)\ln(1-\lambda)]k_B N$$

Taking the zigzag edge at λ=1/4 as an example, our DFT calculations give an estimation of $\Delta H$ as 28.4 meV per edge site. Such a positive value means that phase separation is energetically favorable, which can be easy understood from the edge geometry. The H saturated edge is parallel to the surface while the metal-passivated edge bends to the surface. Therefore, to form a homogenous partially H-saturated edge, there is a notable strain induced energy penalty. At high temperature, such an energy preference is expected to be outperformed by the entropic effect. For λ=1/4, $\Delta S$ is $-4.846 \times 10^{-5}$ eV per edge site per Kelvin. Therefore, a homogeneous configuration becomes thermodynamically more favorable when the temperature is higher than 586.1 K. Since almost all experimental temperatures are higher than this temperature, homogeneous partially H-saturated edge is thus highly relevant in CVD based graphene growth.

To systematically study the edge configuration under different growth conditions, a KMC model is built by considering the equilibrium between edge H atoms and surface H adatoms. As required KMC parameters, energy barriers for H attachment/detachment are predicted from first principles. It turns out that these barriers are sensitive to the local configuration of the edge (Figure 3). For example, if neighboring edge carbon atoms have already been saturated by H, the graphene edge becomes locally less bent to the surface and accordingly H attachment becomes more difficult. Here, we limit the local environment to the nearest neighbors. Test calculations indicate that farther sites only have a limited effect. Since we are now interested in the detailed edge H configuration, the mean field approximation is not used here and all local edge configurations are explicitly considered.

From KMC simulations, the average H coverage of graphene edges is determined under different $H_2$ partial pressures (Figure 4). As expected, graphene edges are metal-passivated under low $H_2$ pressures and H-saturated under high $H_2$ pressures.[41] At the same time, there is a wide range of hydrogen pressures, where graphene edges have a hybrid termination pattern and their edge H coverage increases gradually with the $H_2$ pressure. Compared to zigzag edges, armchair edges are easier to be H-saturated, which can be attributed to the fact that the formation energy of H on armchair edges is lower than that on zigzag edges.[40] More importantly, by checking configurations in the KMC trajectories, we confirm that, in the hybrid termination region, graphene edges are indeed homogeneous without phase separation, which agrees with the conclusion we reached from the estimation of the configuration entropy.

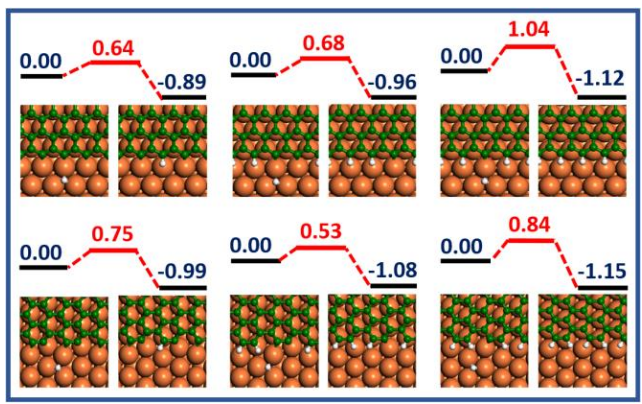

**Figure 3.** Minimum energy paths for H attachment to graphene edges. Red, green, and white balls represent Cu, C, and H, respectively. Energies are in eV.

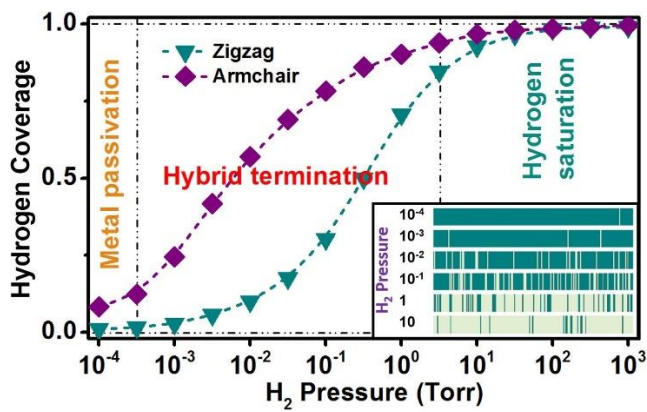

**Figure 4.** Average H coverage of zigzag and armchair graphene edges under different $H_2$ partial pressures. (Inset) Distribution of hydrogen atoms at zigzag graphene edges in a typical KMC snapshot after the equilibrium is reached, where $H_2$ pressures are marked in Torr. Light-green represents H-saturated sites and dark-cyan represents metal-passivated sites.

With the surface concentrations and edge configurations clarified, the next step is determining dynamics of the attachment of C-contained species at graphene edges. In this study, we use zigzag
10

edges, which are more frequently observed under some experimental conditions,[21,42,43] as an example to demonstrate the difference between hydrogen saturated and metal passivated edge configurations. Later we will see that the physical picture obtained here based on zigzag edges are universally applicable to all types of graphene edges. As shown in Figure 5, energy barriers of CH, $CH_2$, $C_2$, and $C_2H$ attachment onto metal-passivated zigzag edge sites are all lower than 0.6 eV, while those of C monomer and $C_2H_2$ are slightly larger than 1 eV. Since it leads to the formation of two C-C bonds, the energy gain from $C_2H_x$ attachment generally is larger than that of $CH_x$.

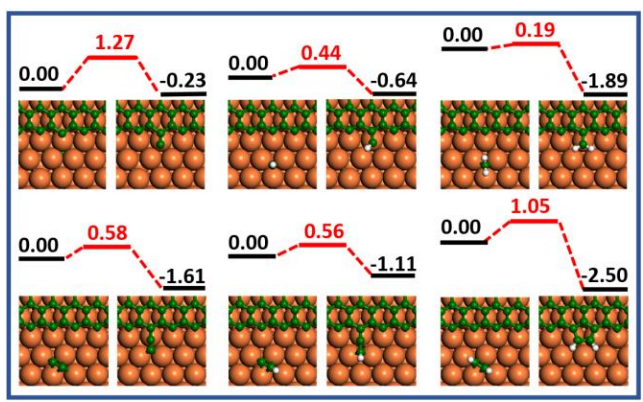

**Figure 5.** Minimum energy paths for C-contained species attachment at metal-passivated zigzag edge sites. Red, green, and white balls represent Cu, C, and H, respectively. Energies are in eV.

As shown in Figure 6, at H-saturated edge sites, attachment of both $C_2$ and CH produces completely $sp^2$ saturated structure, which generally leads to stable final states and low detachment possibility after their attachment. Attachment of CH can be accomplished via two steps. C-C bond is formed in the first step and H on the original edge carbon atom is then transferred to the attached carbon atom. For $C_2$ attachment, there are two possible reaction pathways. After the first C-C bond is formed, the next step can be either formation of the other C-C bond or H-transfer to the attached C atom. Both CH and $C_2$ have a higher attachment barrier at H-saturated edge sites compared to metal-passivated sites. This is a general trend, since C-H bond is stronger than C-Cu bond and H-saturated edge sites are also higher than metal-passivated edge sites. Notice that the same effect will also make the detachment barrier from H-saturated edge sites higher than metal-passivated sites.

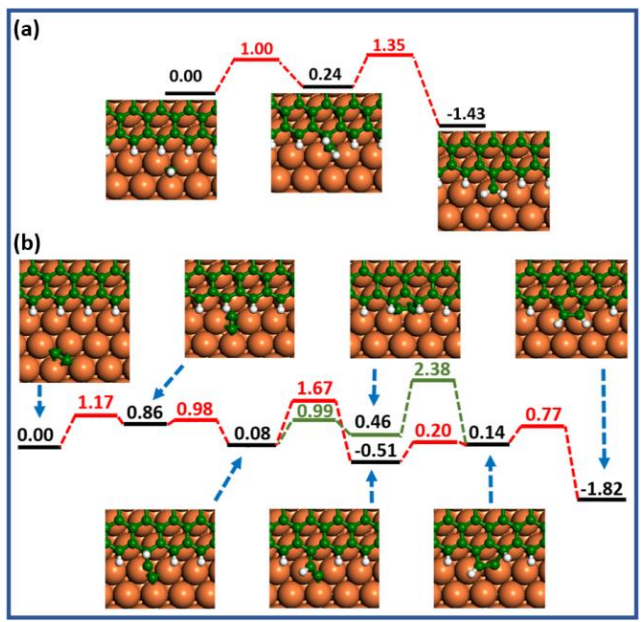

**Figure 6.** Minimum energy paths for the attachment of (a) CH and (b) $C_2$ at H-saturated zigzag edge sites. Red, green, and white balls represent Cu, C, and H, respectively. Energies are in eV.

With kinetic parameters obtained from DFT calculations, KMC simulations are performed to identify the dominant growth pathways in different experimental conditions. The obtained steady-state concentrations of different species during graphene growth are similar to the initial concentrations prior to graphene growth (Figure 2 and Figure S11). To get information about the kinetic pathways during graphene growth, we count the occurring times of different events in KMC trajectories. Figure 7 shows an example under a typical growth condition with the $H_2$ and $CH_4$ partial pressures as 0.1 and 10 Torr, respectively, where metal-passivated edge sites are dominant.

Starting from gas phase $CH_4$, dehydrogenation steps proceed in sequence toward the formation of CH radical. From CH, there are three possible subsequent pathways, i.e. further dehydrogenation, combination to form $C_2H_2$, and graphene edge attachment. Since CH attachment has a relatively low barrier, the last pathway is preferred. However, almost all CH attachment events are followed by a detachment process, as a result of the relatively low detachment barrier of CH. Therefore, CH on the edge is not stable enough to wait for the next attachment event before it is detached and its net contribution to graphene growth is thus small. In contrast, the other two pathways starting from CH are more effective with respect to their reverse reactions. Formation of $C_2H_2$ is strongly exothermic and the gas phase can be considered as a sink of $C_2H_2$. Although CH dehydrogenation is endothermic, the formed C monomer can readily form $C_2$ dimer which is favorable from both the thermodynamic and kinetic points of view.[15] $C_2$ can then be easily attached to graphene edges. Due to its higher detachment barrier and also higher concentration, $C_2$ attachment is much more important for graphene growth than CH attachment. Although the number of $C_2$ attachment events is comparable to CH attachment but the net attached C atoms are much more (1534:115). Therefore, $C_2$ attachment is the dominant pathway for graphene growth at this condition.

When the $H_2$ pressure is increased, concentration of CH can outperform that of $C_2$. More importantly, the graphene edge becomes mainly H-saturated. Unlike at metal-passivated sites, CH attached at an H-saturated edge site is very stable with a detachment barrier even higher than that of $C_2$. As a result, $C_2$ attachment becomes less important than CH attachment under high $H_2$ pressures (Figure 8a). This result is consistent with a recent experimental kinetics analysis,[44] suggesting that CH is the building unit of graphene growth under relatively high $H_2$ pressures. Although the dominant feeding species and kinetic pathways of graphene growth is sensitive to the $H_2$ partial pressure, $CH_4$ pressure shows only a limited effect (Figure 8b). The reason is that H adatom concentration and thus the graphene edge configuration is mainly determined by the former instead of the latter.

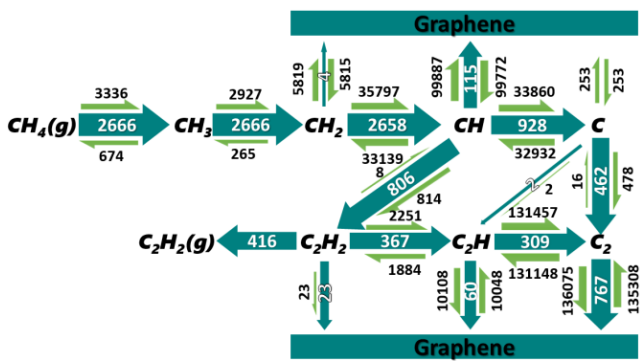

**Figure 7.** Occurring times of different events in a KMC trajectory with the $H_2$ and $CH_4$ pressures to be 0.1 and 10 Torr, respectively. Diffusion on the surface and $H_2$ adsorption/desorption are not shown. Gas phase species are marked with a "g" character in the parenthesis.

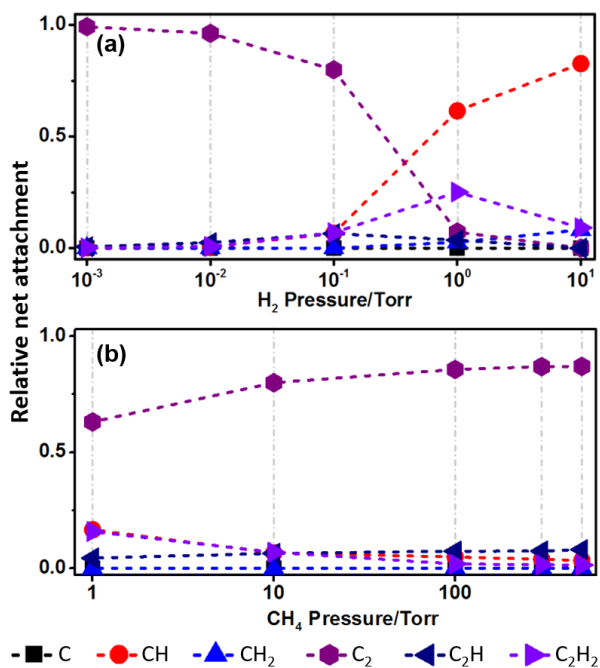

**Figure 8.** Relative net contribution (attachment minus detachment) of each species to graphene growth under different $H_2$ and $CH_4$ pressures. (a) $CH_4$ pressure is fixed to 10 Torr. (b) $H_2$ pressure is fixed at 0.1 Torr. Total net attachment is normalized to be 1.

Our results indicate that, to become a dominant feeding species, the specific surface species should simultaneously have a high surface concentration, a low edge attachment barrier, and also a high detachment barrier. The last condition can easily be overlooked while it is very critical for CVD based graphene growth, where surface concentrations of C-contained species are extremely low (typically below $10^{-5}$ ML) due to the weak interaction between $CH_4$ and the Cu surface. The calculated sticking coefficient of $CH_4$ on Cu(111) at 1300 K is as low as $1.11 \times 10^{-7}$ (see Supporting Information for details). As a result of the low surface carbon concentration, unless there is a high detachment barrier, attached species will always be detached before the arrival of the next carbon species. This is confirmed by the fact that effective attachment numbers are typically several orders of magnitude smaller than the corresponding raw attachment numbers (Figure 7), which also leads to the similarity between the steady-state concentrations of different species during graphene growth and their initial concentrations prior to graphene growth. Only those species being stable enough at graphene edges can be finally survived. Based on our DFT calculations, detachment of C-contained species is generally much easier at metal-passivated edges compared to H-saturated edges, which means that hydrogen saturation stabilizes graphene edges.

Such a picture naturally explains some experimental controversies. For example, there is an debate about if it is possible to grow graphene using the CVD technique but without $H_2$ gas provided.[21,22,45,46] We find that the answer depends on the $CH_4$ pressure. It is possible only if the $CH_4$ pressure is relatively high, which indicates that hydrogen has a catalytic effect on graphene growth but the underlying mechanism is elusive.[21,22,26] Now it becomes clear that such a catalytic effect comes from H-saturation induced edge stabilization. When graphene edges have a low affinity to attached carbon, a higher $CH_4$ partial pressure is required to maintain a relatively high surface concentration of C-contained species and accordingly a short interval time between sequential attachment events. Our simulations indicate that armchair edges are more preferred to be H-saturated under the same $H_2$ pressure, which suggests that armchair edges may grow faster.

The H catalytic effect from edge stabilization will finally be saturated when almost all graphene edge sites become H-saturated. On the other hand, increase of the $H_2$ pressure decreases the concentrations of C-contained species. Therefore, we expect that an unnecessarily high $H_2$ pressure will prohibit graphene growth. This is confirmed by the experimental observation that graphene growth rate first increases then decreases with the $H_2$ partial pressure.[21] The corresponding transition point (0.2~8 Torr) also reasonably agrees with the point predicted by our simulation where graphene edges start to be fully H-saturated (1~10 Torr). Of course, if $CH_4$ input is stopped, $H_2$ will etch grown graphene by helping detached C-contained species to be quickly hydrogenated and desorbed from the surface. Without such a help, there is a higher possibility for detached species to be attached back to a graphene edge, since detachment is an endothermic process. Therefore, the experimentally observed competition between growth and etching[21,44] is in essence a competition between hydrogen induced stabilization of graphene edge and reduction of C-contained species concentration.

Notice that $CH_4$ and $H_2$ partial pressures used in this study may not be able to precisely compare with experiment. For example, a relatively high theoretical $CH_4$ pressures should be adopted to implicitly consider the imperfection of the experimentally used Cu substrate, since defects provide stronger binding and thus a higher adsorption rate of $CH_4$. This effect is less important for $H_2$, which already has a high adsorption rate on a pristine copper surface. At the same time, in ambient pressure (AP) CVD with an inert carrier gas, mass transport from gas phase to the substrate becomes the rate-limiting step,[18,47] which will lead to a lower effective pressure for both $CH_4$ and $H_2$ compared to low pressure (LP) CVD. Considering that the copper substrate is smoother in APCVD,[48,49] its effective pressure will become even lower. This is consistent with the fact that the $CH_4$ and $H_2$ partial pressures used in APCVD are usually higher than those in LPCVD (Figure S19 in Supporting Information).

**Conclusions**

In summary, we have obtained the overall picture of graphene CVD growth on the Cu substrate by a systematical study combining first principles calculations and KMC simulations. Dominant kinetic pathways of graphene growth under different $H_2$ pressures have been identified. Such a detailed kinetic information then leads to the central finding of this study, i.e. the main role hydrogen played in graphene growth is stabilizing the edges of growing graphene islands. These results provide valuable new insights in gaining more precise control of graphene growth or more generally two-dimensional atomic crystal van der Waals epitaxy with the industry compatible CVD technique.

ASSOCIATED CONTENT

**Supporting Information**.

The following files are available free of charge.

Details about KMC simulations, results of test calculations, and summary of experimental conditions of graphene growth. (PDF)

AUTHOR INFORMATION

**Corresponding Author**

*E-mail: zyli@ustc.edu.cn

ACKNOWLEDGMENT

This work was partially supported by the NSFC (21173202, 21421063, and 21573201), the MOST (2016YFA0200604 and 2014CB932700), the CUSF, the NSFC-Guangdong Joint Fund, and by USTC-SCC, SCCAS, Tianjin, and Shanghai Supercomputer Centers.